\documentstyle[aps]{revtex}

\begin{document}
\author{G. Y. Chee}
\address{Physics Department, Liaoning Normal University, Dalian, 116029, China}
\author{Ye Zhang}
\address{China Center of Advanced Science and Technology (World Laboratory), P. O.\\
Box 8730, Beijing, 100080, China and Physics Department, Liaoning Normal\\
University, Dalian, 116029, China}
\author{Yongxin Guo}
\address{Physics Department, Liaoning University, Shenyang , 110036, China}
\title{Gravitational energy-momentum and the Hamiltonian formulation of the
teleparallel gravity}

\begin{abstract}
The transformation properties of the gravitational energy-momentum in the
teleparallel gravity are analyzed. It is proved that the gravitational
energy-momentum in the teleparallel gravity can be expressed in terms of the
Lorentz gauge potential, and therefore is not covariant under local Lorentz
transformations. On the other hand, it can also be expressed in terms of the
translation gauge field strength, and therefore is covariant under general
coordinate transformations. A simplified Hamiltonian formulation of the
teleparallel gravity is given. Its constraint algebra has the same structure
as that of general relativity, which indicates the equivalence between the
teleparallel gravity and general relativity in the Hamiltonian formulation.

\end{abstract}

\maketitle

\section{Introduction}

Recently, as a description of gravity equivalent to general relativity the
teleparallel gravity has attracted renewed attention [1-5] owing to many
salient features of it. First of all, the teleparallel gravity can be
regarded as a translational gauge theory [1, 2, 4, 6], which make it
possible to unify gravity with other kinds of interactions in the gauge
theory framework.. In this direction interesting developments [7] have been
achieved in the context of Ashtekar variables [8]. Another advantage of the
teleparallel gravity concerns energy-momentum, its representation,
positivity and localization [1, 2, 5]. In the context of general relativity
no tensorial expression for the gravitational energy-momentum density can
exist owing to the equivalence principle [9]. On the other hand, because of
its simplicity and transparency the teleparallel gravity seems to be much
more appropriate than general relativity to deal with the problem of the
gravitational energy-momentum. It is proved that [10, 1, 2, 5] in the
teleparallel gravity there exists a gravitational energy-momentum tensor
which is covariant under general coordinate transformations and global
Lorentz transformations. However, it is not covariant under local Lorentz
transformations. The question arises if we can improve it further to obtain
a energy-momentum object which is covariant under general coordinate
transformations as well as local Lorentz transformations. An answer will be
given in this paper.

The teleparallel gravity is characterized by a vanishing curvature and a
nonvanishing torsion. The vanishing of the curvature may be a conceptual
advantage of this formulation in the sense that it may actually define a
background structure.. The identification and establishment of a background
structure is an important issue to be considered since in quantum gravity,
at least from the particle physics point of view, one would ultimately deal
with the energy and momentum of the excitations of the gravitational field,
and these excitations must defined with respect to a background structure.

For a teleparallel geometry there is a preferred class of frames, which
greatly simplify computations. They can be obtained by selecting any frame
at one point and parallel transporting it to all other points. Since the
curvature vanishes, the parallel transport is path independent so the
resultant frame field is globally well defined and then their
transformations are also global, for example, a global Lorentz group. This
means that the vanishing of the curvature or the existence of the absolute
parallel (teleparallel) structure excludes the possibility of the
localization of the global transformation group, e.g. Lorentz group and then
prohibits introducing the corresponding gauge field. In such a teleparallel
frame field\ the connection coefficients (Lorentz gauge fields)\ vanish.
Teleparallel frame fields are unique up to a global (constant , rigid)
linear transformation. In other words, an important feature of the
teleparallel gravity is that the frame field transforms under a global{\it \ 
}Lorentz group. Consequently, the gravitational energy-momentum is a tensor
with respect to coordinate transformations and a global Lorentz group but
not a tensor with respect to a local Lorentz group. A proof of this
conclusion will be given in Sec. II.

Attempts at identifying an energy-momentum density for gravity in the
context of general relativity lead only to various energy-momentum complexes
which are pseudotensors and then a new quasilocal approach which can be
traced back to the early work of Penrose [11] has been proposed and become
widely accepted [5, 12]. According to this approach quasilocal
energy-momentum can be obtained from the Hamiltonian. Every energy-momentum
pseudotensor is associated with a legitimated Hamiltonian boundary term.
Hence, the pseudotensors are quasilocal and acceptable. In the Hamiltonian
formulation of the teleparallel gravity a proof of the positive
gravitational energy was obtained [5].

Concerning the problem of localization of gravitational energy-momentum,
some Hamiltonian formulations of the teleparallel gravity with various gauge
fixing have been presented recently [2, 3, 5]. Furthermore, a consistently
established Hamiltonian formulation not only guarantees field variables to
have a well defined time evolution but also allow us to understand the
physical meaning of the theory from a different perspective. The Hamiltonian
formulation of general relativity reveals the intrinsic structure of the
theory: the time evolution of the field is determined by the scalar and
vector constraints. This is a essential feature for the canonical approach
to the quantum theory of gravity. As is well known, the teleparallel gravity
is equivalent to general relativity, it is naturally to ask if their
Hamiltonian formulations have the same structures. It will be shown this is
the case. In Sec. III a simplified Hamiltonian formulation of the
teleparallel gravity is established and then in Sec. IV its constraint
algebra is obtained under a new gauge fixing. One can find that not only the
Hamiltonian but also the constraint algebra of the teleparallel gravity has
the same structure as that of general relativity, which indicates the
equivalence between teleparallel gravity and general relativity in the
Hamiltonian formulation.

\section{Gravitational energy-momentum and its transformation property}

Usually, it is asserted on the basis of the principle of equivalence that
the gravitational energy cannot be localized [9]. The principle of
equivalence requires that the gravitational field can be made to vanish by a
transformation in a sufficiently small region of the spacetime, which leads
to recognizing the connection on the spacetime manifold as the strength of a
gravitational field. However, if the gravitational energy-momentum density
consists of the curvature rather than the connection like the case in
electromagnetism, we would not have the problem of energy localization.

As is well known, a physical object has different transformation characters
under different transformation groups. Therefore the answer to the question
about the transformation characters of the gravitational energy-momentum
object depends on the choice of variables, the choice of the gauge group and
the expression of the energy-momentum object itself. For example, it depends
on whether the expression of the energy-momentum object consists of the
gauge potential or the gauge field strength. In a gauge theory the gauge
potential is not covariant under the corresponding gauge group. As a result,
the self current of the gauge field derived from Noether theorems is not
covariant naturally [13].

In this section we will see that the gravitational energy-momentum can be
expressed in terms of the Lorentz gauge potential, therefore it is not
covariant under local Lorentz transformations. On the other hand it can also
be expressed in terms of the translation gauge field strength and therefore
is covariant under general coordinate transformations.

We start with a common relation between the tetrad $e^I\!_\mu $, the spin
connection $\omega _\mu \!^I\!_J$, and the affine connection $\Gamma ^\rho
\!_{\nu \mu }\;$[14] 
\begin{equation}
\partial _\mu e^{I{}}\!_\nu +\omega _\mu \!^I\!_Je^J\!_\nu -\Gamma ^\rho
\!_{\mu \nu }e^I\!_\rho =0,
\end{equation}
where $I,J,\ldots =0,1,2,3$ are the internal indices and $\mu ,\nu ,\ldots
=0,1,2,3\;$are the spacetime indices. If we define the Cartan connection
[1,15] 
\begin{equation}
\Gamma _{(c)\mu \nu }^\rho =e\!_I\!^\rho \partial _\mu e^{I{}}\!_\nu ,
\end{equation}
then (1) leads to 
\begin{eqnarray}
\Gamma _{(c)\mu \nu }^\rho &=&\Gamma ^\rho \!_{\mu \nu }-\omega _\mu
\!^I\!_Je_I\!^\rho e^J\!_\nu  \nonumber \\
&=&\Gamma ^\rho \!_{\mu \nu }-\omega _\mu \!^\rho \!_\nu  \nonumber \\
&=&\{_\mu \!^\rho \!_\nu \}+K^\rho \!_{\mu \nu }-\omega _\mu \!^\rho \!_\nu ,
\end{eqnarray}
where 
\begin{equation}
\omega _\mu \!^\rho \!_\nu =\omega _\mu \!^I\!_Je_I\!^\rho e^J\!_\nu ,
\end{equation}
and $\{_\mu \!^\rho \!_\nu \}$,$\;K^\rho \!_{\mu \nu }$ is the Christoffel
connection and the affine contorsion, respectively. By introducing the
Cartan torsion 
\begin{equation}
T_{(c)\mu \nu }^\rho =\Gamma _{(c)\mu \nu }^\rho -\Gamma _{(c)\nu \mu }^\rho
,
\end{equation}
and the Cartan contorsion [1,15] 
\begin{equation}
K_{(c)}^\rho \!_{\mu \nu }=\frac 12(T_{(c)}^\rho \!_{\nu \mu }+T_{(c)\mu
}\!^\rho \!_\nu +T_{(c)\nu }\!^\rho \!_\mu ),
\end{equation}
we can obtain from (3) 
\begin{equation}
T_{(c)\mu \nu }^\rho =T^\rho \!_{\mu \nu }-\omega _\mu \!^\rho \!_\nu
+\omega _\nu \!^\rho \!_\mu ,
\end{equation}
and 
\begin{equation}
K_{(c)}^\rho \!_{\mu \nu }=K^\rho \!_{\mu \nu }+\omega _{\!\mu \nu }\!^\rho .
\end{equation}
where 
\begin{equation}
T^\rho \!_{\mu \nu }=2\Gamma ^\rho \!_{[\mu \nu ]}=\Gamma ^\rho \!_{\mu \nu
}-\Gamma ^\rho \!_{\nu \mu },
\end{equation}
is the affine torsion and 
\begin{equation}
K^\rho \!_{\mu \nu }=\frac 12(T^\rho \!_{\mu \nu }+T_\mu \!^\rho \!_\nu
+T_\nu \!^\rho \!_\mu ),
\end{equation}
is the affine contorsion [16].

In [14] the Cartan torsion $T_{(c)JK}^I=T_{(c)\mu \nu }^\rho e^I\!_\rho
e_J\!^\mu e_K\!^\nu $ with a factor $\frac 12$ is called the anholonomity
and denoted as $C_{JK}\!^I$, i.e. 
\begin{equation}
C_{JK}\!^I=\frac 12T_{(c)JK}^I=\frac 12T_{(c)\mu \nu }^\rho e^I\!_\rho
e_J\!^\mu e_K\!^\nu
\end{equation}
and then is given a different geometric meaning. It is not a gauge-covariant
object. In the 'holonomic gauge' $C_{JK}\!^I$ vanishes and then we have a
natural (or coordinate ) coframe. In this case (7) gives the relation
between the anholonomity, the affine torsion and the spin connection: 
\begin{equation}
2C_{\mu \nu }\!^\rho =2C_{JK}\!^Ie_I\!^\rho e^J\!_\mu e^K\!_\nu =T^\rho
\!_{\mu \nu }-\omega _\mu \!^\rho \!_\nu +\omega _\nu \!^\rho \!_\mu .
\end{equation}

In the case of vanishing affine torsion 
\begin{equation}
T^\rho \!_{\mu \nu }=0,
\end{equation}
as in the usually general relativity, we have 
\begin{equation}
K^\rho \!_{\mu \nu }=0,
\end{equation}
and then 
\begin{equation}
K_{(c)}^\rho \!_{\mu \nu }=\omega _{\mu \!\nu }\!^\rho .
\end{equation}
As a result (3) reads 
\begin{equation}
\Gamma _{(c)\mu \nu }^\rho =\{_\mu \!^\rho \!_\nu \}+K_{(c)}^\rho \!_{\mu
\nu },
\end{equation}
and we are led to the theories given by Hayashi, Shirafuji [15], de Andrade,
Guillen and Pereira [1] which are equivalent to general relativity and
called theories of teleparallel gravity. In these theories the curvature of
the Cartan connection vanishes: 
\begin{equation}
R_{(c)}^\rho \!_{\sigma \mu \nu }=\partial _\mu \Gamma _{(c)\sigma \nu
}^\rho -\partial _\nu \Gamma _{(c)\sigma \mu }^\rho +\Gamma _{(c)\tau \mu
}^\rho \Gamma _{(c)\sigma \nu }^\tau -\Gamma _{(c)\tau \nu }^\rho \Gamma
_{(c)\sigma \mu }^\tau \equiv 0,
\end{equation}
while the curvature of the Christoffel connection 
\begin{equation}
R^\rho \!_{\sigma \mu \nu }=\partial _\mu \{_\sigma \!^\rho \!_\nu
\}-\partial _\nu \{_\sigma \!^\rho \!_\mu \}+\{_\tau \!^\rho \!_\mu
\}\{_\sigma \!^\tau \!_\nu \}-\{_\tau \!^\rho \!_\nu \}\{_\sigma \!^\tau
\!_\mu \}
\end{equation}
does not. For these theories one can say that the spacetime is a Weitzenbock
spacetime with respect to the Cartan connection or a Riemann spacetime with
respect to the Christoffel connection.

Maluf develops another kind of teleparallel description of general
relativity [2] in which the curvature of the affine connection vanishes
while the affine torsion does not. Therefore one can say that the spacetime
of the Maluf's description is a Weitzenbock spacetime with respect to the
affine connection.

According to [1], the Lagrangian of the gravitational field can be chosen as

\begin{equation}
{\cal L}_G=-\frac{^{(4)}ec^4}{16\pi G}S^{\rho \mu \nu }T_{(c)\rho \mu \nu },
\end{equation}
where $^{(4)}e=\det (e^I\!_\mu )$, and 
\begin{equation}
S^{\rho \mu \nu }=\frac 12(K_{(c)}^{\mu \nu \rho }-g^{\rho \nu
}T_{(c)}^{\sigma \mu }\!_\sigma +g^{\rho \mu }T_{(c)}^{\sigma \nu }\!_\sigma
).
\end{equation}
The energy-momentum density of the gravitational field is: 
\begin{equation}
^{(4)}ej_I\!{}^\rho =-\frac{^{(4)}ec^4}{4\pi G}e_I\!^\lambda S_\mu \!^{\nu
\rho }T_{(c)}^\mu \!_{\nu \lambda }-e_I\!^\rho {\cal L}_G.
\end{equation}
The quantity $j_I\!{}^\rho $ transforms covariantly under a general
spacetime coordinate transformation, and is invariant under local
translation of the tangent-space coordinates. However, it transform
covariantly only under a global Lorentz transformation. How does it behave
under a local Lorentz transformation ? The answer is given in the following..

From (7) and (13) we can obtain 
\begin{equation}
T_{(c)\mu \nu }^\rho =\omega _\nu \!^\rho \!_\mu -\omega _\mu \!^\rho \!_\nu
,
\end{equation}
and 
\begin{equation}
S_\mu \!^{\nu \rho }=\frac 12(\omega ^\rho \!_\mu \!^\nu -\delta _\mu ^\rho
\omega _\sigma \!^{\sigma \nu }+\delta _\mu ^\nu \omega _\sigma \!^{\sigma
\rho }),
\end{equation}
which lead to 
\begin{equation}
S^{\rho \mu \nu }T_{(c)\rho \mu \nu }=\omega _\rho \!^{\mu \nu }\omega _\mu
\!^\rho \!_\nu -\omega _\rho \!^{\rho \nu }\omega _\mu \!^\mu \!_\nu ,
\end{equation}
and 
\begin{equation}
S_\mu \!^{\nu \rho }T_{(c)}^\mu \!_{\nu \lambda }=\frac 12\omega ^\rho
\!_\mu \!^\nu (\omega _\lambda \!^\mu \!_\nu -\omega _\nu \!^\mu \!_\lambda
)-\frac 12\omega _\mu \!^{\mu \nu }(\omega _\lambda \!^\rho \!_\nu -\omega
_\nu \!^\rho \!_\lambda )-\frac 12\omega _\mu \!^{\mu \rho }\omega _\nu
\!^\nu \!_\lambda .
\end{equation}
The equations (21), (19), (24) and (25) indicate that $j_I\!^\rho $ consists
of the Lorentz gauge potential $\omega _\rho \!^\mu \!_\nu $ algebraically
and then is neither covariant nor invariant under local Lorentz
transformations. If local Lorentz transformations are not introduced in the
theory then $\omega _\rho \!^\mu \!_\nu =0$ which leads to $j_I\!^\rho =0$.
Therefore, we are led to the conclusion that in teleparallel gravity the
gravitational energy-momentum is not covariant under local Lorentz
transformations because it consists of the Lorentz gauge potential.

Using (15), (6) and (5) we can compute 
\begin{eqnarray}
\omega _{\mu \nu \rho } &=&K_{(c)\rho \mu \nu }=\frac 12(T_{(c)\mu \rho \nu
}+T_{(c)\nu \rho \mu }+T_{(c)\rho \mu \nu })  \nonumber \\
&=&(e_{I\rho }\partial _{[\nu }e^I\!_{\mu ]}+e_{I\mu }\partial _{[\rho
}e^I\!_{\nu ]}+e_{I\nu }\partial _{[\rho }e^I\!_{\mu ]}),
\end{eqnarray}
\begin{equation}
\omega ^{\nu \mu \rho }=g^{\nu \tau }g^{\rho \sigma }e_I\!^\mu \partial
_{[\sigma }e^I\!_{\tau ]}+g^{\mu \lambda }g^{\rho \sigma }e_I\!^\nu \partial
_{[\sigma }e^I\!_{\lambda ]}+g^{\mu \lambda }g^{\nu \tau }e_I\!^\rho
\partial _{[\tau }e^I\!_{\lambda ]},
\end{equation}
\begin{eqnarray}
S^{\rho \mu \nu }T_{(c)\rho \mu \nu } &=&\omega _{\mu \nu \rho }\omega ^{\nu
\mu \rho }-\omega _\mu \!^{\mu \rho }\omega _\nu \!^\nu \!_\rho  \nonumber \\
&=&\frac 14T_{(c)\rho \mu \nu }T_{(c)}^{\rho \mu \nu }+\frac 12T_{(c)\rho
\mu \nu }T_{(c)}^{\nu \mu \rho }+T_{(c)\rho \mu }\!\!^\rho T_{(c)}^{\nu
\!}\!_\nu \!^\mu
\end{eqnarray}
Using the translation gauge field strength 
\begin{equation}
F^I\!_{\mu \nu }=c^2e^{I\!}\!_\rho T_{(c)\mu \nu }^\rho =c^2(\partial _\mu
e^I\!_\nu -\partial _\nu e^I\!_\mu ),
\end{equation}
the Eq. (19) can be written as 
\begin{equation}
{\cal L}_G=\frac{^{(4)}e}{64\pi G}(\eta _{IJ}g^{\mu \lambda }g^{\nu \tau
}+2e_I\!^\tau e_J\!^\nu g^{\mu \lambda }-4e_I\!^\mu e_J\!^\lambda g^{\nu
\tau })F^I\!_{\mu \nu }F^J\!_{\lambda \tau .}
\end{equation}
The first term of (21) is 
\begin{eqnarray}
\frac{^{(4)}ec^4}{4\pi G}e_I\!^\lambda S_\mu \!^{\nu \rho }T_{(c)}^\mu
\!_{\nu \lambda } &=&\frac{^{(4)}e}{8\pi G}e_I\!^\tau [\frac 12%
(e_J{}^\lambda e_K{}^\mu g^{\rho \nu }+e_J{}^\rho e_K{}^\mu g^{\lambda \nu
}-\eta _{JK}g^{\lambda \mu }g^{\rho \nu })  \nonumber \\
&&-e_J{}^\nu e_K{}^\rho g^{\lambda \mu }+e_J{}^\nu e_K{}^\lambda g^{\rho \mu
}]F^J\!_{\mu \nu }F^K\!_{\lambda \tau .}
\end{eqnarray}
From (20), (30) and (31) we can see that the current $j_I{}^\rho $ consists
of the translation gauge field strength $F^I\!_{\mu \nu }$ and then is
covariant under local translations. This makes teleparallel gravity
different from the usual gauge field theories where the Noether current of
the gauge fields is not covariant under corresponding gauge transformations.

\section{3+1 decomposition}

In order to obtain the Hamiltonian formulation of the theory a foliation in
the spacetime manifold $M$ should be introduced. Assuming that $M=\Sigma
\times R$ for some space-like Manifold $\Sigma $, we can choose a time
function $t$ with nowhere vanishing gradient $(dt)_\mu $ such that each $%
t=const$ surface $\Sigma _t$ is diffeomorphic to $\Sigma $. Introduce a time
flow vector $t^\mu $ satisfying $t^\mu (dt)_\mu =1$, we can decompose it
perpendicular and parallel to $\Sigma _t$: $t^\mu =Nn^\mu +N^\mu $, where $%
n^\mu $ is the time-like normal at each point of $\Sigma _t$ and $N$, $N^\mu 
$ are the lapse function and the shift vector , respectively. The spacetime
metric $g_{\mu \nu }$ introduces a spatial metric $q_{\mu \nu }$ on each $%
\Sigma _t\;$by the formula 
\begin{equation}
q_{\mu \nu }=g_{\mu \nu }+n_\mu n_\nu ,
\end{equation}
a triad 
\begin{equation}
e^I\!_i=q_i^\mu e^I\!_\mu
\end{equation}
defined on $\Sigma _t$ with $e=\det (e^I\!_i)=^{(4)}e/N$ and a volume
element $\epsilon _{ijk}$ of $q_{ij}$. Since the affine torsion vanishes by
(13), from this section we will drop the subscript $(c)$ of the Cartan
torsion $T_{(c)\mu \nu }^I$ and simply denote it by $T^I\!_{\mu \nu }$. Then
the Lagrangian

\begin{equation}
{\cal L}_G=\frac{^{(4)}ec^4}{64\pi G}(\eta _{IJ}g^{\mu \lambda }g^{\nu \tau
}+2^{(4)}e_I\!^{\tau (4)}e_J\!^\nu g^{\mu \lambda }-4^{(4)}e_I\!^{\mu
(4)}e_J\!^\lambda g^{\nu \tau })T^I\!_{\mu \nu }T^J\!_{\lambda \tau }
\end{equation}
can be rewritten as

\begin{eqnarray}
{\cal L}_G &=&\frac{Nec^4}{64\pi G}%
(T_{ijk}T^{ijk}+2T_{ikj}T^{jki}-4T^i{}_{ki}T^{jk}{}_j-T^{\perp
}{}_{ij}T^{\perp ij}+4T^{\perp i}{}_jT^j{}_{\perp i}  \nonumber \\
&&-8T^{ji}{}_jT^{\perp }{}_{\perp i}-2T_{j\perp }{}^iT^j{}_{\perp
i}-2T^i{}_{\perp j}T_{j\perp }{}^i+4T^i{}_{\perp i}T^j{}_{\perp j}),
\end{eqnarray}
here we have used the notation 
\[
V_{\perp }=n^\mu V_{\mu ,}, 
\]
and 
\[
V^{\perp }=n_\mu V^\mu . 
\]
Only the factors 
\begin{equation}
T^I{}_{\perp i}=n^\mu T^I{}_{\mu i}=\frac 1N(\stackrel{\cdot }{e^I{}_i}%
-\partial _ie^I{}_0+N^jT^I{}_{ij})
\end{equation}
contain the time derivatives $\stackrel{\cdot }{e^I{}_i}={\cal L}_te^I{}_i$,
the canonical momentum conjugate to $e^I{}_i$ is

\begin{equation}
\widetilde{p}_I{}^i=\frac{\partial {\cal L}_G}{\partial \stackrel{\cdot }{%
e^I{}_i}}=\frac{ec^4}{16\pi G}(T^{\perp i}{}_I-2n_IT^{ji}{}_j-T_{I\perp
}{}^i-T^i{}_{\perp I}+2e_I{}^iT^j{}_{\perp j}),
\end{equation}
which has the properties

\begin{equation}
\widetilde{p}^{(ij)}=\widetilde{p}_I{}^{(i}e^{I|j)}=\frac{ec^4}{8\pi G}%
(q^{ij}T^k{}_{\perp k}-T^{(i}{}_{\perp }{}^{j)}),
\end{equation}
\begin{equation}
\widetilde{p}^{[ij]}=\widetilde{p}_I{}^{[i}e^{I|j]}=\frac{ec^4}{16\pi G}%
T^{\perp ij},
\end{equation}
and 
\begin{equation}
\widetilde{p}_{\perp }{}^i=n^I\widetilde{p}_I{}^i=\frac{ec^4}{8\pi G}%
T^{ji}{}_j.
\end{equation}
Using these results and 
\[
\widetilde{p}_I{}^i\stackrel{\cdot }{e^I{}_i}=NT^I{}_{\perp i}\widetilde{p}%
_I{}^i+N^jT^I{}_{ji}\widetilde{p}_I{}^i+\widetilde{p}_I{}^i\partial
_ie^I{}_0, 
\]
\[
\widetilde{p}_I{}^i\partial _ie^I{}_0=\partial _i(\widetilde{p}%
_I{}^ie^I{}_0)-Nn^I\partial _i\widetilde{p}_I{}^i-N^je^I{}_j\partial _i%
\widetilde{p}_I{}^i, 
\]
\begin{eqnarray*}
NT^I{}_{\perp i}\widetilde{p}_I{}^i &=&\frac{Nec^4}{16\pi G}(T^j{}_{\perp
i}T^{\perp i}{}_j-2T^{\perp }{}_{\perp i}T^{ji}{}_j-T^j{}_{\perp i}T_{j\perp
}{}^i \\
&&-T^j{}_{\perp i}T^i{}_{\perp j}+2T^i{}_{\perp i}T^j{}_{\perp j}),
\end{eqnarray*}
we can obtain the canonical Hamiltonian density 
\begin{eqnarray}
{\cal H}_G &=&\widetilde{p}_I{}^i\stackrel{\cdot }{e^I{}_i}-{\cal L}_G 
\nonumber \\
&=&N{\cal H}_{\perp }+N^j{\cal H}_j+\partial _i\widetilde{B}{}^i,
\end{eqnarray}
where 
\begin{eqnarray}
{\cal H}_{\perp } &=&\frac{2\pi G}{ec^4}(\widetilde{p}^2-2\widetilde{p}%
^{(ij)}\widetilde{p}_{(ij)})-n^I\partial _i\widetilde{p}_I{}^i  \nonumber \\
&&-\frac{ec^4}{64\pi G}%
(T_{ijk}T^{ijk}+2T_{ikj}T^{jki}-4T^i{}_{ki}T^{jk}{}_j-T^{\perp
}{}_{ij}T^{\perp ij})  \nonumber \\
&=&\frac{2\pi G}{ec^4}(\widetilde{p}^2-2\widetilde{p}^{ij}\widetilde{p}%
_{ji}-2\widetilde{p}{}_{\perp }{}^i\widetilde{p}_{\perp i})-\frac{ec^4}{%
64\pi G}(T_{ijk}T^{ijk}+2T_{ikj}T^{jki})-n^I\partial _i\widetilde{p}_I{}^i,
\end{eqnarray}
\begin{equation}
{\cal H}_j=T^I{}_{ji}\widetilde{p}_I{}^i-e^I{}_j\partial _i\widetilde{p}%
_I{}^i=\widetilde{p}_I{}^i\partial _je^I{}_i-\partial _i\widetilde{p}_j{}^i,
\end{equation}
\begin{equation}
\widetilde{B}{}^i=\widetilde{p}_I{}^ie^I{}_0=t^\mu e^I{}_\mu \widetilde{p}%
_I{}^i,
\end{equation}
with $\widetilde{p}=e^I{}_i\widetilde{p}_I{}^i$. The term $\partial _i%
\widetilde{B}{}^i$ in ${\cal H}_G$ is essential for the quasilocal
energy-momentum of the gravitational field [5] but it can be ignored in the
constraint analysis.

The primary constraints are 
\begin{equation}
\phi _N=\widetilde{p}_N=\frac{\partial {\cal L}_G}{\partial \stackrel{\cdot 
}{N}}=0,
\end{equation}
\begin{equation}
\phi _i=\widetilde{p}_i=\frac{\partial {\cal L}_G}{\partial \stackrel{\cdot 
}{N^i}}=0,
\end{equation}
and the primary Hamiltonian density is 
\begin{equation}
{\cal H}_p={\cal H}_G+\alpha \widetilde{p}_N+\beta ^i\widetilde{p}_i.
\end{equation}
For the further constraint analysis we need to compute 
\begin{eqnarray}
\frac{\delta {\cal H}_{\perp }}{\delta e^I{}_i} &=&\frac{c^4}{16\pi G}%
\partial _j[e(T_I{}^{ji}+2T^{[ij]}{}_I)]  \nonumber \\
&&+\frac{2\pi G}{ec^4}[e_I{}^i(\widetilde{p}^2-2\widetilde{p}^{jk}\widetilde{%
p}_{kj}-2\widetilde{p}{}_{\perp }{}^j\widetilde{p}_{\perp j})+2\widetilde{p}%
\widetilde{p}_I{}^i-4\widetilde{p}_I{}^j\widetilde{p}_j{}^i-4e_{Ij}%
\widetilde{p}{}_{\perp }{}^j\widetilde{p}{}_{\perp }{}^i]  \nonumber \\
&&-\frac{ec^4}{64\pi G}%
[e_I{}^i(T_{ijk}T^{ijk}+2T_{ikj}T^{jki})+4(T_{jkI}T^{jki}+T_{jkI}T^{ikj}+T_{jkI}T^{kji})],
\end{eqnarray}
\begin{equation}
\frac{\delta {\cal H}_{\perp }}{\delta \widetilde{p}_I{}^i}=\frac{4\pi G}{%
ec^4}(\widetilde{p}e^I{}_i-\widetilde{p}^I{}_i-\widetilde{p}_i{}^I)+\partial
_in^I,
\end{equation}
\begin{equation}
\frac{\delta {\cal H}_j}{\delta e^I{}_i}=-\partial _j\widetilde{p}_I{}^i,
\end{equation}
and 
\begin{equation}
\frac{\delta {\cal H}_j}{\delta \widetilde{p}_I{}^i}=\partial _je^I{}_i.
\end{equation}
The consistency conditions 
\begin{equation}
\stackrel{\cdot }{\phi _N}=\{\phi _N,H_p\}=-\frac{\delta H_p}{\delta N}=-%
{\cal H}_{\perp }=0,
\end{equation}
\begin{equation}
\stackrel{\cdot }{\phi _N}=\{\phi _i,H_p\}=-\frac{\delta H_p}{\delta N^i}=-%
{\cal H}_i=0.
\end{equation}
lead two secondary constraints, while the conditions 
\begin{eqnarray*}
\stackrel{\cdot }{{\cal H}_{\perp }} &=&\{{\cal H}_{\perp },H_p\}=0, \\
\stackrel{\cdot }{{\cal H}_i} &=&\{{\cal H}_i,H_p\}=0,
\end{eqnarray*}
are only some conditions imposed on the Lagrange multipliers $N,N^i,\alpha
,\beta ^i\;$ and lead to no new constraints.

Thus by following the Dirac constraint analysis we find that the phase space
($\Gamma _{TG},\Omega _{TG}$) of the teleparallel gravity is coordinatized
by the pair ($e^I{}_i,\widetilde{p}_I{}^i$) and has symplectic structure 
\begin{equation}
\Omega _{TG}=\int_\Sigma d\widetilde{p}_I{}^i\wedge de^I{}_i.
\end{equation}
Ignoring the surface integral the Hamiltonian is given by 
\begin{equation}
H_{TG}=\int_\Sigma N{\cal H}_{\perp }+N^j{\cal H}_j.
\end{equation}

\section{Constraint algebra}

In order to obtain the constraint algebra we construct the constraint
functions by smearing ${\cal H}_{\perp }$\ and ${\cal H}_i$\ with test
fields $N$\ and $N^i$\ on $\Sigma $ following the approach of Ashtekar and
Romano[8,17]: 
\begin{equation}
C(N)=\int_\Sigma N{\cal H}_{\perp },
\end{equation}

\begin{equation}
C(\overrightarrow{N})=\int_\Sigma N^i{\cal H}_i.
\end{equation}
Under the gauge condition 
\begin{equation}
\partial _iN^i=0,
\end{equation}
we have 
\begin{equation}
\frac{\delta C(\overrightarrow{N})}{\delta e^I{}_i}=-{\cal L}_{%
\overrightarrow{N}}\widetilde{p}_I{}^i,
\end{equation}
and 
\begin{equation}
\frac{\delta C(\overrightarrow{N})}{\delta \widetilde{p}_I{}^i}={\cal L}_{%
\overrightarrow{N}}e^I{}_i.
\end{equation}
The Hamiltonian vector field of $C(\overrightarrow{N})$ generates a
transformation with the parameter $\varepsilon ,$ the changes of $e^I{}_i$
and $\widetilde{p}_I{}^i$ under this transformation are, respectively, 
\begin{eqnarray}
\delta e^I{}_i &=&-\{C(\overrightarrow{N}),e^I{}_i\}\varepsilon =\frac{%
\delta C(\overrightarrow{N})}{\delta \widetilde{p}_I{}^i}\varepsilon
=\varepsilon {\cal L}_{\overrightarrow{N}}e^I{}_i, \\
\delta \widetilde{p}_I{}^i &=&-\{C(\overrightarrow{N}),\widetilde{p}%
_I{}^i\}\varepsilon =-\frac{\delta C(\overrightarrow{N})}{\delta e^I{}_i}%
\varepsilon =\varepsilon {\cal L}_{\overrightarrow{N}}^I{}\widetilde{p}%
_I{}^i,
\end{eqnarray}
which means that the Hamiltonian vector field of $C(\overrightarrow{N})$ on
the phase space is the lift of the vector field $N^i$ on $\Sigma $. One can
say that the vector constraint $C(\overrightarrow{N})$ generates a space
translation along the shift vector $\overrightarrow{N}$ on $\Sigma $.

Using the Hamiltonian equations 
\begin{equation}
\stackrel{\cdot }{e^I{}_i}=\frac{\delta H_p}{\delta \widetilde{p}_I{}^i}=%
\frac{\delta C(N)}{\delta \widetilde{p}_I{}^i}+\frac{\delta C(%
\overrightarrow{N})}{\delta \widetilde{p}_I{}^i},
\end{equation}
and 
\[
\stackrel{\cdot }{\widetilde{p}_I{}^i}=-\frac{\delta H_p}{\delta e^I{}_i}=-%
\frac{\delta C(N)}{\delta e^I{}_i}-\frac{\delta C(\overrightarrow{N})}{%
\delta e^I{}_i}, 
\]
in the case $N^i=0$, we have 
\begin{equation}
\frac{\delta C(N)}{\delta e^I{}_i}=\stackrel{\cdot }{-\widetilde{p}_I{}^i}=-%
{\cal L}_{\overrightarrow{t}}\widetilde{p}_I{}^i,
\end{equation}
and 
\begin{equation}
\frac{\delta C(N)}{\delta \widetilde{p}_I{}^i}=\stackrel{\cdot }{e^I{}_i}=%
{\cal L}_{\overrightarrow{t}}e^I{}_i.
\end{equation}
The changes of $e^I{}_i$ and $\widetilde{p}_I{}^i$ under the diffeomorphism
with the parameter $\varepsilon $ generated by the Hamiltonian vector field
of $C(N)$ are, respectively,

\begin{eqnarray}
\delta e^I{}_i &=&-\{C(N),e^I{}_i\}\varepsilon =\frac{\delta C(N)}{\delta 
\widetilde{p}_I{}^i}\varepsilon =\varepsilon {\cal L}_{\overrightarrow{t}%
}e^I{}_i, \\
\delta \widetilde{p}_I{}^i &=&-\{C(N),\widetilde{p}_I{}^i\}\varepsilon =-%
\frac{\delta C(N)}{\delta e^I{}_i}\varepsilon =\varepsilon {\cal L}_{%
\overrightarrow{t}}^I{}\widetilde{p}_I{}^i,
\end{eqnarray}
which means that when restricted to the constrained phase space the
Hamiltonian vector field of $C(N)$ is the lift of the vector field $t^\mu
=Nn^\mu $ on $\Sigma $. In other words the scalar constraint $C(N)$
generates a lapse of time.

In order to calculate the Poisson brackets of $C(N)$\ and $C(N)$, we derive
two formulas.\ If $f(M)$ is any real-valued function on the phase space of
the form 
\begin{equation}
f(M)=\int_\Sigma M^{a\cdots b}{}_{c\cdots d}\widetilde{f}_{a\cdots
b}{}^{c\cdots d}(e^I{}_i,\widetilde{p}_I{}^i),
\end{equation}
where $M^{a\cdots b}{}_{c\cdots d}=M^{a\cdots b}{}_{c\cdots d}(%
\overrightarrow{x})$ is any tensor field independent of $e^I{}_i$ and$\;%
\widetilde{p}_I{}^i$ on $\Sigma $ then 
\begin{eqnarray*}
\{C(\overrightarrow{N)},f(M)\} &=&\int_\Sigma \frac{\delta C(\overrightarrow{%
N})}{\delta e^I{}_i}\frac{\delta f(M)}{\delta \widetilde{p}_I{}^i}-\frac{%
\delta C(\overrightarrow{N})}{\delta \widetilde{p}_I{}^i}\frac{\delta f(M)}{%
\delta e^I{}_i} \\
&=&-\int_\Sigma {\cal L}_{\overrightarrow{N}}\widetilde{p}_I{}^i\frac{\delta
f(M)}{\delta \widetilde{p}_I{}^i}+{\cal L}_{\overrightarrow{N}}e^I{}_i\frac{%
\delta f(M)}{\delta e^I{}_i} \\
&=&-\int_\Sigma M^{a\cdots b}{}_{c\cdots d}{\cal L}_{\overrightarrow{N}}%
\widetilde{f}_{a\cdots b}{}^{c\cdots d}.
\end{eqnarray*}
Integrating by parts and throwing away the surface integral, we obtain 
\begin{equation}
\{C(\overrightarrow{N)},f(M)\}=f({\cal L}_{\overrightarrow{N}}M).
\end{equation}
Let $C(\overrightarrow{M)},\;C(M)$ to be $f(M)$ we have 
\begin{equation}
\{C(\overrightarrow{N)},C(\overrightarrow{M})\}=C({\cal L}_{\overrightarrow{N%
}}\overrightarrow{M})=C([\overrightarrow{N},\overrightarrow{M}]).
\end{equation}
\begin{equation}
\{C(\overrightarrow{N)},C(M)\}=C({\cal L}_{\overrightarrow{N}}M).
\end{equation}
By the similar way we can get for $C(N)$: 
\begin{equation}
\{C(N),f(M)\}=-{\cal L}_{\overrightarrow{t}}f(M)+f({\cal L}_{\overrightarrow{%
t}}M)-f({\cal L}_{\overrightarrow{N}}M).
\end{equation}
Let $f(M)=C(M),$we have 
\begin{equation}
\{C(N),C(M)\}=C({\cal L}_{\overrightarrow{t}}M)-C({\cal L}_{\overrightarrow{t%
}}M),
\end{equation}
owing to the consistency condition $\stackrel{\cdot }{C(M)}\;={\cal L}_{%
\overrightarrow{t}}C(M)=0$.

The equations (70), (71) and (73) indicate that the constraint algebra is
closed and the constraints $C(N)$ and $C(\overrightarrow{N)}$ are first
class, which is very similar to the case in general relativity. The
equations (61), (62), (66) and (67) mean that the first class constraints $%
C(N)$ and $C(\overrightarrow{N)}$ generate the corresponding gauge
transformations, the spacetime translations. We have shown that under the
gauge condition $\partial _iN^i=0$ the constraint algebra of the
teleparallel gravity has the same structure as that of general relativity.

\section{Conclusion}

From a common relation between the tetrad $e^I\!_\mu $, the spin connection $%
\omega _\mu \!^I\!_J$, and the affine connection $\Gamma ^\rho \!_{\nu \mu }$%
, it is proved that the gravitational energy-momentum in the teleparallel
gravity can be expressed in terms of the Lorentz gauge potential as well as
the translation gauge field strength. It is this characteristic that leads
to the complicated transformation property of the gravitational
energy-momentum: it is not covariant under local Lorentz transformations but
is covariant under general coordinate transformations. The lack of a local
Lorentz covariance can be considered as the teleparallel manifestation of
the pseudotensor character of the gravitational energy-momentum in general
relativity. It is not possible to define a local Lorentz covariant gauge
current in the teleparallel gravity, consequently the corresponding
gravitational energy-momentum in general relativity can not be represented
by a true tensor.. Therefore the apparent covariance of the gravitational
energy-momentum density is actually cosmetic. The quasilocal approach is the
farthest we can go in the direction to deal with the problem of
gravitational energy-momentum in the framework of gauge theories. The
Hamiltonian formulation of the teleparallel gravity is the same as the
Hamiltonian formulation of general relativity. Under the gauge condition $%
\partial _iN^i=0$ the two constraints $C(N)$ and $C(\overrightarrow{N)}$ are
first class and generate the corresponding gauge transformations, the
spacetime translations. The constraint algebra of the teleparallel gravity
has the same structure as that of general relativity. Thus we have shown
that the teleparallel gravity is equivalent to general relativity not only
in the Lagrangian formulation but also in the Hamiltonian formulation
although their geometries are different. In microscopic physics the
teleparallel description is more useful because of its flat background
structure.

{\bf ACKNOWLEDGMENT}$\;$

This work is supported by the National Science Foundation Grants of China
No.19875023 and the Science Foundation Grants of the Education Department of
Liaoning Province No. 20041011.

\end{document}